\documentclass[twocolumn,prb,amssymb,superscriptaddress,showpacs]{revtex4}
\usepackage{graphicx}

\begin{document}

\title{Giant suppression of the Drude conductivity
due to quantum interference in disordered two-dimensional systems}

\author{G.~M.~Minkov}
\affiliation{Institute of Metal Physics RAS, 620219 Ekaterinburg,
Russia}
\author{A.~V.~Germanenko}
\author{O.~E.~Rut}
\affiliation{Institute of Physics and Applied Mathematics, Ural
State University, 620083 Ekaterinburg, Russia}

\author{A.~A.~Sherstobitov}
\affiliation{Institute of Metal Physics RAS, 620219 Ekaterinburg,
Russia}
\author{B.~N.~Zvonkov}
\address{Physical-Technical Research Institute, University of
Nizhni Novgorod, 603600 Nizhni Novgorod, Russia}

\date{\today}

\begin{abstract}
Temperature and magnetic field dependences of the conductivity in
heavily doped, strongly disordered two-dimensional quantum well
structures GaAs/In$_x$Ga$_{1-x}$As/GaAs are investigated within wide
conductivity and temperature ranges. Role of the interference in the
electron transport is studied in the regimes when the phase breaking
length $L_\phi$ crosses over the localization length $\xi\sim
l\exp{(\pi k_Fl/2)}$ with lowering temperature, where $k_F$ and $l$
are the Fermi quasimomentum and mean free path, respectively. It has
been shown that all the experimental data can be understood within
framework of simple model of the conductivity over delocalized
states. This model differs from the conventional model of the weak
localization developed for $k_Fl\gg 1$ and $L_\phi\ll\xi$ by one
point: the value of the quantum interference contribution to the
conductivity is restricted not only by the phase breaking length
$L_\phi$ but by the localization length $\xi$ as well. We show that
just the quantity
$(\tau_\phi^\ast)^{-1}=\tau_\phi^{-1}+\tau_\xi^{-1}$ rather than
$\tau_\phi^{-1}$, where $\tau_\phi\propto T^{-1}$ is the dephasing
time and $\tau_\xi\sim\tau\exp(\pi k_F l)$, is responsible for the
temperature and magnetic field dependences of the conductivity over
the wide range of temperature and disorder strength down to the
conductivity of order  $10^{-2}\, e^2/h$.

\end{abstract}

 \maketitle

\section{Introduction}
\label{sec:int}

It is commonly accepted  that all states  are localized in
two-dimension (2D) systems at arbitrary disorder.\cite{Abr79} The
localization length  at zero magnetic field can be estimated
as\cite{Lee85}
\begin{equation}
\xi\simeq l \exp(\pi k_Fl/2), \label{eq00}
\end{equation}
where $l$ is the mean free path and $k_F$ is the Fermi
quasimomentum. This result was obtained for noninteracting electrons
at zero temperature. Another characteristic length which appears at
finite temperature is a phase breaking length, $L_\phi$. Just the
ratio between these characteristic lengths determines the
conductivity regime.

When $\xi\gg L_\phi$ one can consider electrons as delocalized ones.
The conductivity, $\sigma$, in this regime  can be well described
starting with the classical Drude model. When the quantum effects
are ignored, the conductivity is given by the Drude formula
\begin{equation}
\sigma_{0}=e^2n\tau/m=\pi G_0 k_F l, \label{eq10}
\end{equation}
where $G_0=e^2/(2\pi^2 \hbar)$, $n$ is the electron density, $\tau$
and $m$ are the transport relaxation time and electron effective
mass, respectively. Two quantum corrections to the conductivity
determine both the temperature and magnetic field dependences of the
conductivity at low enough temperature when the electron gas is
degenerated and phonon scattering does not influence the momentum
relaxation. They are the weak localization (WL) or interference
correction and the correction caused by the electron-electron ({\it
e-e}) interaction. The WL correction  is negative and
logarithmically increases in magnitude with the decreasing
temperature when the spin relaxation is absent. The interaction
correction, as a rule,  is also negative and increases in absolute
value with the lowering temperature if the gas parameter $r_s$ is
less than unity and the system is in the diffusion regime, $T\tau\ll
1$ (hereafter we set $\hbar=1$, $k_B=1$). Thus, the conductivity at
$\xi\gg L_\phi$ decreases with $T$-decrease and remains
significantly higher than $G_0$.

In opposite case, $\xi\ll L_\phi$,  the electrons can be considered
as well localized. The conductivity in this regime is hopping and
should be described in  framework of the percolation
theory.\cite{Shkl84} The conductivity under this condition is
significantly less than $G_0$.

Thus, the range of the conductivity values, where both approaches
are wrong in the strict sense, is very wide. For example, an unusual
metallic-like temperature dependence of the conductivity is
experimentally observed just within this range.\cite{Abr01,Alt01} A
crossover from the weak to strong localization or, by other words,
the crossover from conductivity over delocalized states to the
conductivity over localized states takes place also here.

In order to avoid possible misunderstanding we would like to note at
the very beginning that throughout this paper we will use terms
``conductivity over {\it delocalized} states'' and ``conductivity
over {\it localized} states'' in the following senses. The
conductivity over {\it delocalized} states means that one can
describe the transport phenomena (the temperature, magnetic- and
electric-field dependences of the longitudinal and transverse
conductivity) starting from the model of free electrons. The
incoherent and coherent scattering and the {\em e-e} interaction are
taken into account  as perturbation. Just for this case the
quantitative expressions for temperature and magnetic field
dependences of the conductivity was
obtained.\cite{Hik80,Kaw84,Wit87,Knap96,Zdu97,Dmit97}

The term ``conductivity over {\it localized} states'' means that one
has to describe the transport phenomena starting from the model of
well localized states, considering the transitions between them as
perturbation.  This approach holds when the ``conductivity'' between
two nodes  in the equivalent Miller-Abrahams network is
significantly less than the conductivity quantum $e^2/h$. Taking
into account the disorder, the conductivity of the whole network is
relatively small, $\sigma\ll e^2/h$.

Experimentally, the conductivity of 2D systems at $\sigma\lesssim
G_0$ was studied in the number of
papers.\cite{Keus97-1,Keus97,Bud98,Shlim00,Gersh00,Min02,Camin03} In
most cases the  data are interpreted from the position of the
hopping mechanism of the conductivity. The expressions obtained in
framework of the percolation theory are widely used for quantitative
analysis. Only in Refs.~\onlinecite{Bud98} and \onlinecite{Min02} it
is pointed out that some of the dependences  look like that for the
conductivity over delocalized states.

As  mentioned above,  the crossover from weak to strong localization
is governed by the ratio $\xi/L_\phi$. That is why, it would be
interesting to trace the changing of the temperature and magnetic
field dependences of the conductivity and Hall effect at continuous
variation of the $\xi$ to $L_\phi$ ratio. There are two different
ways for that. It can be done by varying either the disorder
strength (i.e., $\xi$) or the temperature (i.e., $L_\phi$). The
first can be realized on structures of different design. For
example, centrally doped quantum well structures, the structures
with remote doping layers, the structures of different doping level,
or the gated structures  can be investigated. However, the electron
density and mobility change simultaneously in this way.

The second way, the changing of the temperature, seems more suitable
because all the parameters remain constant except $L_\phi$. Strongly
disordered (with small $\tau$-value) 2D system with electron gas of
high density is appropriate object in this case. Strong disorder
ensures that the interference correction, which is proportional to
$\ln(\tau_\phi/\tau)$, is comparable in magnitude with the Drude
conductivity at $k_Fl>1$ at accessibly low temperatures. The high
electron density and low mobility guarantee the degeneracy of the
electron gas and negligibility of the phonon scattering up to the
relatively high temperature. Thus, changing the temperature it is
possible  to cross over from the regime $L_\phi>\xi$ to the regime
$L_\phi<\xi$ keeping disorder unchanged. Just this line of attack is
used in Refs.~\onlinecite{Gersh97} and \onlinecite{Khavin98} to
study the crossover from weak to strong localization in
quasi-one-dimensional structures.

In this paper, we present the results of experimental study of the
temperature and magnetic field dependences of the conductivity and
Hall effect in the heavily $\delta$-doped quantum well
GaAs/In$_x$Ga$_{1-x}$As/GaAs structure within wide temperature
range.

\section{samples}
We have investigated two heterostructures, 4261 and 4262, which are
distinguished by doping level. They were grown by metal-organic
vapor-phase epitaxy on a semiinsulating GaAs substrate and consist
of $0.5$-$\mu$m-thick undoped GaAs epilayer, a 80-\AA\
In$_{0.2}$Ga$_{0.8}$As quantum well with Sn $\delta$ layer situated
in the well center and a 2000-\AA\ cap layer of undoped GaAs. The
tin density, $N_d$, in $\delta$ layer was about $2\times
10^{12}$~cm$^{-2}$ and $1\times 10^{12}$~cm$^{-2}$ for structure
4261 and 4262, respectively. Several samples were mesa etched from
each wafer into standard Hall bars and then an Al gate electrode was
deposited onto the cap layer by thermal evaporation. The
measurements were performed in the temperature range $0.4-80$ K at
magnetic field $B$ up to 6 T. The electron density at $V_g=0$ was
$1.8\times10^{12}$ cm$^{-2}$ in the structure 4261 and
$0.9\times10^{12}$ cm$^{-2}$ in the structure  4262. Since  the
results were mostly analogous for both structures, we restrict our
attention in this paper to the structure 4261 only.

\section{Results and discussion}
\label{sec:res}

\begin{figure}[b]
\includegraphics[width=0.8\linewidth,clip=true]{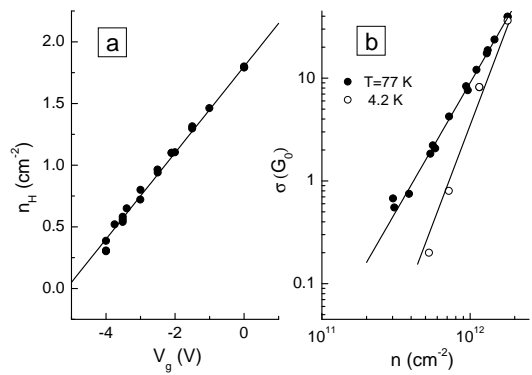}
\caption{ (a) -- The gate-voltage dependence of the Hall density
$n_H=1/(eR_H)$ for $T=77$~K. Symbols are the experimental results.
Straight line is drown with the slope $dn/dV_g=3.5\times
10^{11}$~cm$^{-2}$V$^{-1}$. (b) -- The conductivity in zero magnetic
field for two temperatures as a function of electron density.
Symbols are the experimental results. Lines are provided as a guide
to the eye. }
 \label{F1}
\end{figure}

To characterize the structures investigated, Fig.~\ref{F1}(a) shows
the gate voltage dependence of the Hall density, $n_H=1/(eR_H)$,
measured at $T=77$~K when the interaction correction to the Hall
coefficient is negligible.  One can see that the electron density
linearly depends on the gate voltage with the slope
$dn_H/dV_g=3.5\times 10^{11}$~cm$^{-2}$V$^{-1}$ which accords well
with the structure design and capacitance measurements.
Fig.~\ref{F1}(b) shows the electron-density dependences of the
conductivity at $T=77$~K and $T=4.2$~K. Already this figure shows
that the  change of the conductivity with temperature is strong for
the relatively high electron density. So the conductivity difference
for these temperatures is about one order of magnitude for $n\simeq
5\times 10^{11}$ cm$^{-2}$ (the Fermi energy, $E_F$, is about $20$
meV).

To clarify the role of the quantum corrections at low conductivity
when $k_F l \simeq 1.5-2$ let us analyze the experimental data
starting from the case of $k_F l \gg 1$ when the conventional
theories of the quantum corrections work beyond question. We will
not present all data and restrict ourself by the cases of the key
$k_Fl$-values: (i) the minimal $k_Fl$-value when we do not observe a
deviation from the conventional theories of the quantum corrections
yet; (ii) the case of  the maximal $k_Fl$-value where deviations are
already evident; (iii) the minimal $k_Fl$-value when we are able to
understand all the dependences after some modification of the
conventional theory which takes into account the partial
localization of the electron wave function and, at last, (iv) the
case when a new approach for description of the conductivity is
required  from our point of view.

\begin{figure}
\includegraphics[width=0.8\linewidth,clip=true]{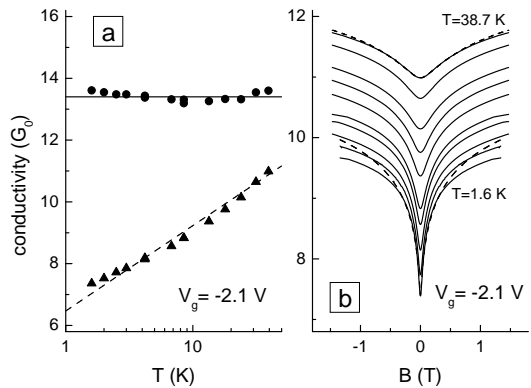}
\caption{(a) -- The temperature dependence of the conductivity at
$B=0$ for $V_g=-2.1$~V.  The values measured experimentally are
shown by triangles. Circles are the Drude conductivity found
experimentally for the different temperatures as described in text.
Solid line is $\sigma_0=13.4\, G_0$. Dashed line is the dependence
$1.2\ln{T}$. (b) -- The magnetic-field dependence of the
conductivity measured for different temperatures: $T= 38.7$, $31.0$,
$23.5$, $17.7$, $13.2$, $8.5$, $6.8$, $4.2$, $2.5$, and $1.6$~K
(from top to the bottom). Solid lines are experimental, dashed lines
are example of the fit by Eq.~(\ref{eq35})  carried out for
$T=1.6$~K and $38.7$~K within the magnetic field range from $-0.3\,
B_{tr}$ to $0.3\, B_{tr}$, $B_{tr}\simeq 1.25$~T.}
 \label{F2}
\end{figure}

Let us begin with the analysis of the data obtained at $V_g=-2.1$~V,
$n=1.1\times10^{12}$ cm$^{-2}$ that corresponds to the first case.
Fig.~\ref{F2}(a) shows the temperature dependence of the
conductivity at $B=0$. One can see that  the $\sigma$-vs-$T$
dependence is close to the logarithmic one over the whole
temperature range. The low-magnetic-field dependences of the
conductivity $\sigma=\rho_{xx}^{-1}$  measured in the temperature
range  from $1.6$~K to $40$~K  are shown in Fig.~\ref{F2}(b). As
evident the magnetoresistance demonstrates the typical features
corresponding to the suppression of weak localization by  magnetic
field: the  temperature decrease leads to the sharpening of the
minima at $B=0$.

To treat the data quantitatively one needs to know the values of the
Drude conductivity $\sigma_0$. As seen from Fig.~\ref{F2}(a) the
change of the conductivity with the temperature increase  for such
disordered systems is strong already for relatively high
conductivity. The conductivity  is enhanced in magnitude more than
fifty percent in our temperature range. Therefore, there are
difficulties in determination of $\sigma_0$. We use the procedure
analogous to that described in Refs.~\onlinecite{Min02} and
\onlinecite{Min04-2}. It allows us to obtain simultaneously the
values both of $\sigma_0$ and of $\tau_\phi$.

The conductivity in the absence of a magnetic field can be written
as
\begin{equation}
\sigma(T)=\sigma_0+ \delta\sigma^{WL}(T) +
\delta\sigma_b^{WL}(T)+\delta\sigma_{ee}(T). \label{eq20}
\end{equation}
Here, $\delta\sigma^{WL}(T)$ and $\delta\sigma^{WL}_b(T)$ stand
for the backscattering\cite{AA85} and
non-backscattering\cite{Dmit97} parts of the interference quantum
correction, respectively. In the diffusion regime, $\tau\ll
\tau_\phi$, they are
\begin{equation}
\delta\sigma^{WL}(T)=\beta\, G_0
\ln\left[\frac{\tau}{\tau_\phi(T)}\right],\,\,\,
\delta\sigma^{WL}_b= G_0\,\ln\,2, \label{eq25}
\end{equation}
where  $\beta$ is equal to unity.\cite{Min04-2}  The last term in
Eq.~(\ref{eq20}) is the quantum correction caused by the {\it e-e}
interaction. For $T\tau\ll 1$, it
reads\cite{Fin83re,Cast84-1,Cast84-2,Cast98}
\begin{eqnarray}
\delta\sigma_{ee}(T)&=&\left\{1+3\left[1-
\frac{\ln(1+F_0^\sigma)}{F_0^\sigma}\right]\right\} G_0\,\ln{T\tau} \nonumber \\
&\equiv& K_{ee}\,G_0\,\ln{T\tau}. \label{eq30}
\end{eqnarray}
where $F_0^\sigma$ is the Fermi-liquid constant. The interaction
correction has been  studied for analogous quantum-well structure in
our previous papers.\cite{Min02,Min03} In particular, it has been
observed experimentally\cite{Min03} that $K_{ee}$ starts to decrease
below $k_Fl\sim 4$ and apparently tends to zero with further
lowering $k_Fl$. Close behavior is demonstrated in the structures
investigated in this paper; $K_{ee}$ practically vanishes when $k_F
l\lesssim 2$. The theoretical explanation of this fact may possible
require going to the next order in disorder in Finkel'stein's
renormalization-group scheme.\cite{Fin83re}  Presenting the values
of $K_{ee}$ for each concrete case, we will not discuss this issue
in this paper anymore.

The shape of the low-field negative magnetoconductance
$\Delta\sigma(B)=\rho_{xx}^{-1}(B)-\rho^{-1}(0)$, caused by the
suppressing  interference correction is  described by the
Hikami-Larkin-Nagaoka (HLN) expression\cite{Hik80,Wit87}
\begin{eqnarray}
\Delta\sigma(B)&=&\alpha\, G_0\, {\cal H}\left(\frac{\tau}{\tau_\phi},\frac{B}{B_{tr}}\right), \nonumber \\
{\cal H}(x,y)& = & \psi\left(\frac{1}{2}+\frac{x}{y}\right) -
\psi\left(\frac{1}{2}+\frac{1}{y}\right)- \ln{x}, \label{eq35}
\end{eqnarray}
where $B_{tr}=\hbar/(2el^2)$ is the transport magnetic field,
$\psi(x)$ is a digamma function, and $\alpha$ is the prefactor. With
taking into account two-loop correction and interplay of weak
localization and interaction, the prefactor $\alpha$ should depend
on the conductivity as follows\cite{Alei99,Min04-2}
\begin{equation}
\alpha=1-\frac{2\,G_0}{\sigma}, \,\,\, \sigma>2.\label{eq40}
\end{equation}
Expressions from (\ref{eq20}) to  (\ref{eq35}) have been used to
find $\sigma_0$ and $\tau_\phi$.  We used a successive approximation
method as in Ref.~\onlinecite{Min02}. For the first approximation we
have set $\sigma_0$ equal to the experimental value $\sigma$ at
$B=0$, found seed values $\tau$ and $B_{tr}$ using $n$ obtained from
the Hall measurements. Then, we have determined $\tau_\phi/\tau$ and
$\alpha$ from the fit of experimental curve of magnetoconductance by
the HLN-expression, Eq.~(\ref{eq35}). After that we have substituted
the ratio $\tau_\phi/\tau$ into Eq.~(\ref{eq25}) and using
experimental value of $K_{ee}$, Eqs.~(\ref{eq30}) and (\ref{eq20})
found the corrected value of $\sigma_0$ (and, consequently, $\tau$
and $B_{tr}$). Found value of $\sigma_0$ was used for next step.  A
convergence of the procedure is good enough. It is sufficient to
make from five to seven iterations to achieve an accuracy of several
percent in the determination of $\sigma_0$ .

The values of $\sigma_0$ found in such a way for different
temperatures are shown in Fig.~\ref{F2}(a) by solid circles. One can
see that they are close to each other. This attests that (i) the
model used is adequate, (ii) the value $\sigma_0$ found by this way
is good estimate for the Drude conductivity. Thus, the Drude
conductivity for the case of $V_g=-2.1$~V is equal to
$\sigma_0=(13.4\pm 0.4)\,G_0$, that corresponds to $k_F l=4.3\pm
0.1$.

\begin{figure}
\includegraphics[width=0.8\linewidth,clip=true]{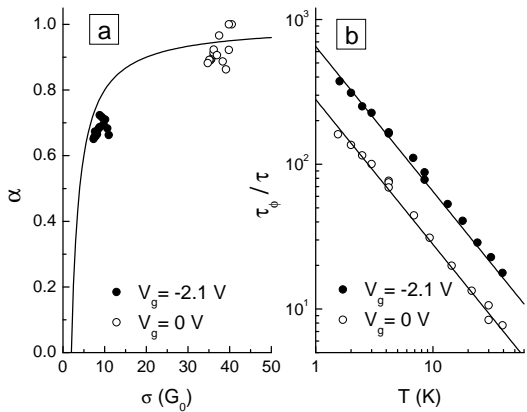}
\caption{(a) -- The prefactor $\alpha$ as a function of conductivity
varied with the temperature for $V_g=-2.1$~V and $0$~V. Symbols are
the data, solid line is Eq.~(\ref{eq40}). (b) -- The temperature
dependence of the $\tau_\phi$ to $\tau$ ratio found experimentally
from the shape of the magnetoconductance curve for $V_g=-2.1$~V and
$0$~V. Solid lines are the $T^{-1}$-law.} \label{F3}
\end{figure}

The parameters $\alpha$ and $\tau_\phi/\tau$ found from the
processing of the magnetoconductivity curves are shown in
Figs.~\ref{F3}(a) and \ref{F3}(b), respectively. For completeness
sake the results for $V_g=0$~V are presented also. Note, the points
in Fig.~\ref{F3}(a) for each gate voltage relate to the different
temperatures. The parameter $K_{ee}$ used for the estimate of
interaction correction is equal here to $0.2$ and $0.25$ for
$V_g=-2.1$~V and $0$~V, respectively. As seen from Fig.~\ref{F3}(a)
the conductivity dependence of the prefactor $\alpha$ is really
close to the theoretical one, Eq.~(\ref{eq40}). The temperature
dependence of the $\tau_\phi$ to $\tau$ ratio  is also close to that
predicted theoretically,\cite{AA85} $\tau_\phi/\tau\propto 1/T$
[Fig.~\ref{F3}(b)].

As seen from  Fig.~\ref{F2}(a) the temperature dependence of the
conductivity at $B=0$ accords with this model too. It is close to
logarithmic with the slope $1+0.2$  where $1$ and $0.2$ come from WL
and interaction corrections, respectively [dashed line in
Fig.~\ref{F2}(a)].

Thus, all the temperature and magnetic field dependences of the
conductivity for large enough $k_Fl$ value are in  quantitative
agreement with the conventional theories of the quantum corrections,
which take into account the two-loop interference correction and
interplay between  interference and interaction. We would like to
emphasize here two important issues. First, the WL correction  is
dominant. For example, at $T=1.6$ K
$\delta\sigma^{WL}+\delta\sigma_b^{WL}\simeq -5.3\,G_0$ while
$\delta\sigma^{ee}\simeq -0.9\, G_0$. Second, both corrections for
such a strongly disordered systems significantly suppress the Drude
conductivity  already at large enough value of $k_Fl$, $k_Fl\simeq
4$.

\begin{figure}[t]
\includegraphics[width=0.8\linewidth,clip=true]{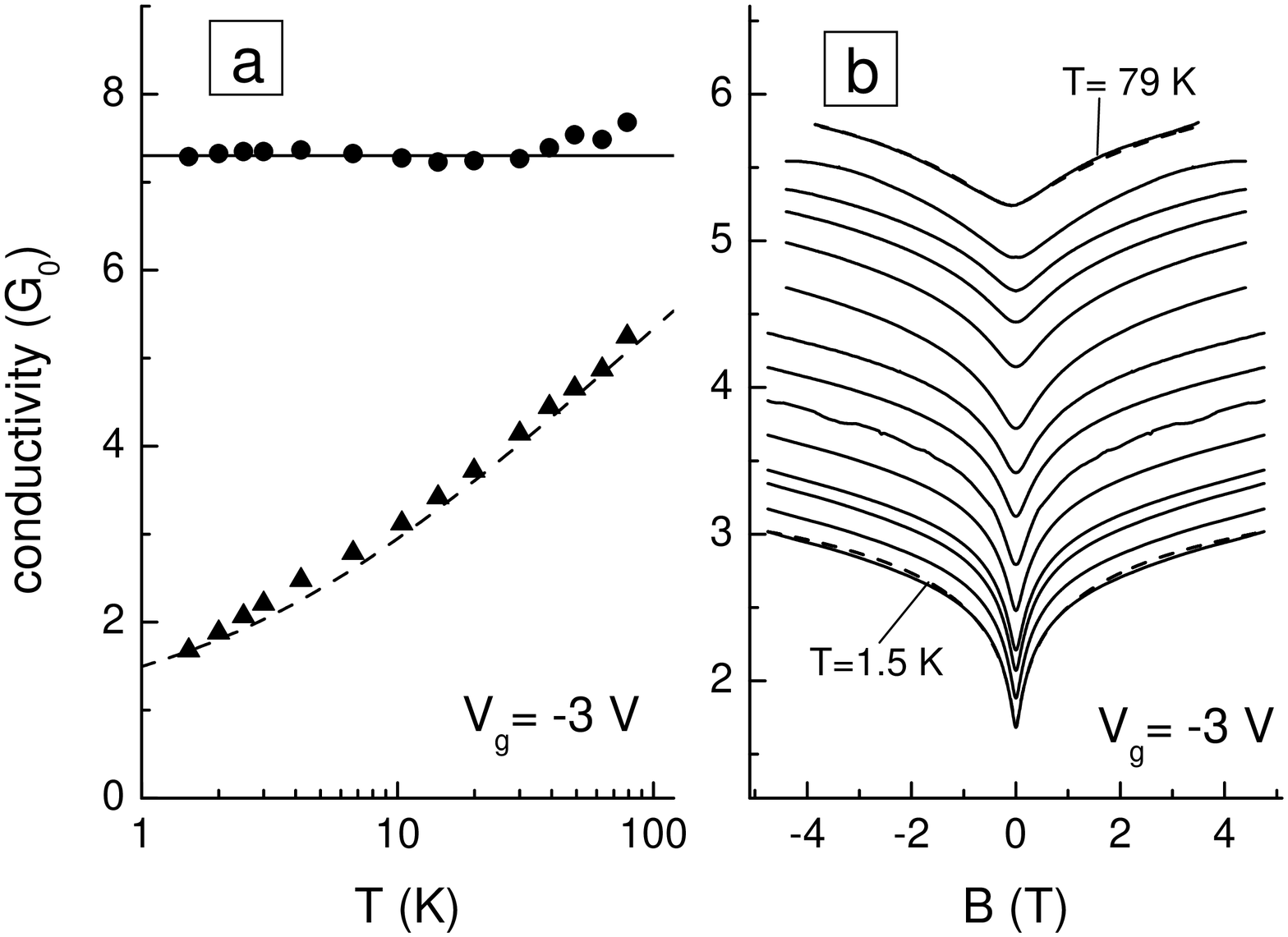}
\caption{(a) -- The $T$-dependence of the conductivity at $B=0$
measured for $V_g=-3$~V. The values measured experimentally shown as
triangles. Circles are the Drude conductivity found experimentally
for the different temperatures as described in text. The solid line
is $\sigma_0=7.3\, G_0$, the dashed line is Eq.~(\ref{eq20})
calculated with $\tau_\phi^\ast/\tau$ shown in Fig.~\ref{F5}(b), the
dotted line the dependence $1.15\ln{T}$. (b) -- The magnetic-field
dependence of the conductivity for different temperatures: $T=79$,
$63$, $49.2$, $39.2$, $30$, $19.9$, $14.4$, $10.4$, $6.7$, $4.2$,
$3.0$, $2.5$, $2.0$, and $1.5$~K (from top to the bottom). Solid
lines are experimental, dashed lines are the fit by Eq.~(\ref{eq35})
carried out within magnetic field range from $-0.3\, B_{tr}$ to
$0.3\, B_{tr}$, $B_{tr}\simeq 3.1$~T.}
 \label{F4}
\end{figure}

\begin{figure}
\includegraphics[width=\linewidth,clip=true]{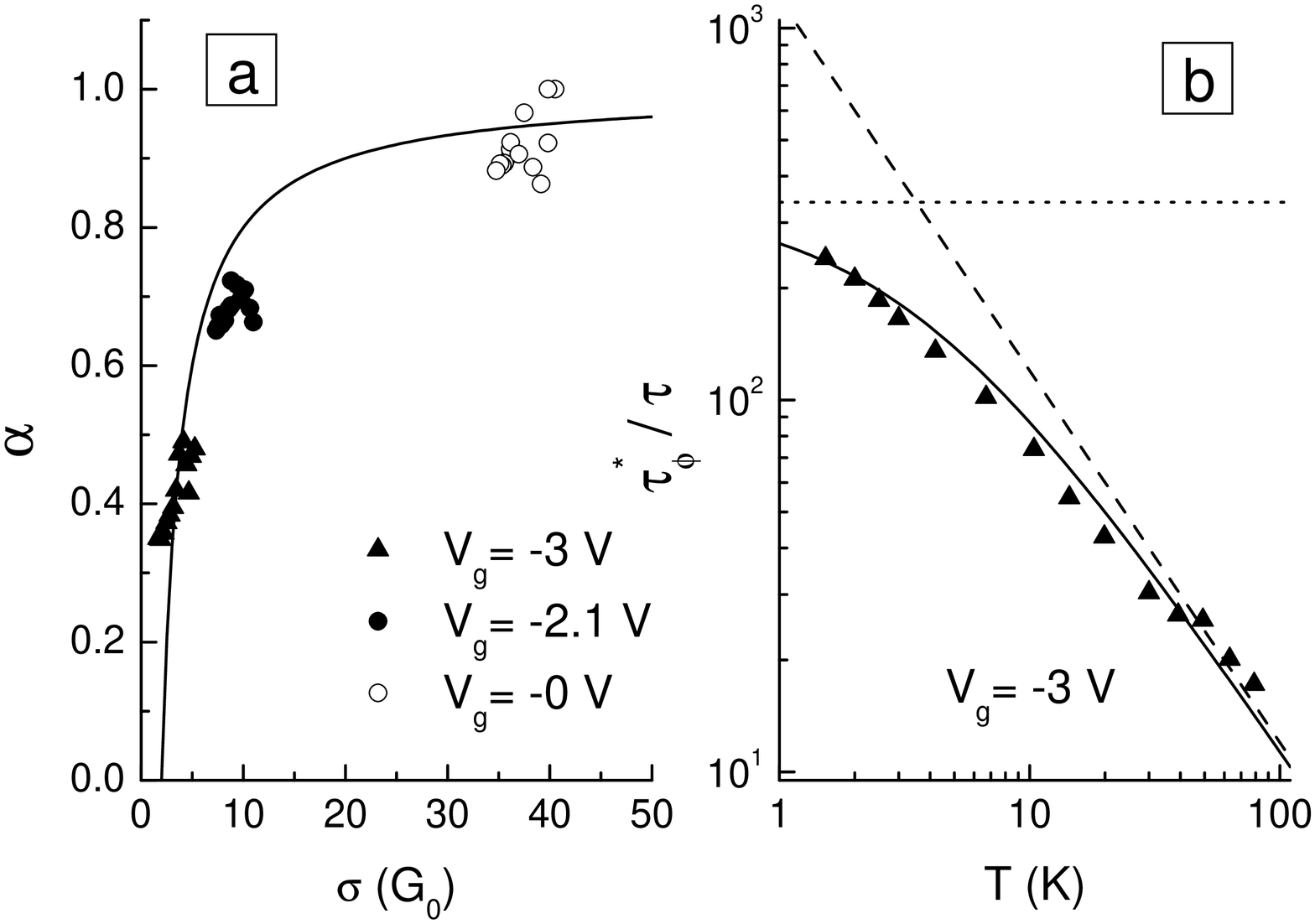}
\caption{(a) -- The prefactor $\alpha$ as a function of the
conductivity for the different  gate voltages. Symbols are the data,
solid line is Eq.~(\ref{eq40}). (b) -- The temperature dependence of
the $\tau^\ast_\phi$ to $\tau$ ratio found experimentally for
$V_g=-3$~V. The dashed line is  $\tau_\phi/\tau=1200/T$, the dotted
line is $\tau_\xi/\tau=340$, the solid line is
$\tau^\ast_\phi/\tau=1/(\tau/\tau_\phi+\tau/\tau_\xi)$. } \label{F5}
\end{figure}

Let us  consider the results for stronger disorder. The temperature
and magnetic field dependences of the conductivity at $V_g=-3$~V
when $n=8\times 10^{11}$~cm$^{-2}$  are presented in Fig.~\ref{F4}.
Again, the negative magnetoresistance is observed over the whole
temperature range up to $T\simeq 80$~K. However, the variation of
the conductivity with temperature  remains close to
$(1+0.15)\,\ln(T)$ (where $0.15$ comes from interaction)  only at
$T>10$~K and slackens noticeably at lower temperature. As well as in
the previous case the magnetoresistance curves are well described by
the HLN expression. The conductivity dependence of the prefactor
$\alpha$ and the temperature dependences of the fitting parameter
$\tau_\phi^\ast$ are presented in Figs.~\ref{F5}(a) and \ref{F5}(b),
respectively (we asterisk here $\tau_\phi$ because it is unknown in
advance how the fitting parameter in the HLN-expression relates to
the phase breaking time at low conductivity, $\sigma<e^2/h$). One
can see that $\alpha$ decreases with the lowering conductivity and
remains close to the theoretical dependence  down to $\sigma\simeq
2.5\,G_0$. The deviation at lower $\sigma$ is not surprising because
the theoretical dependence, Eq.~(\ref{eq40}), takes into account the
first order in $1/\sigma$-expansion only. The temperature dependence
of $\tau_\phi^\ast$ is close to theoretical $1/T$- dependence only
at $T>20$ K and shows the tendency to saturation at lower
temperature.

Before to consider  possible reasons of the deviations of
$\tau_\phi^\ast$-vs-$T$ and $\sigma(B=0)$-vs-$T$ dependences from
that predicted by the conventional theories, let us evaluate the
value of the Drude conductivity by the same way as in previous case.
Using  $\tau_\phi^\ast$ as $\tau_\phi$ in Eq.~(\ref{eq25}) and
$K_{ee}=0.15$ in Eq.~(\ref{eq30}),  we obtain the results presented
in Fig.~\ref{F4}(a) by solid cirles. It is evident that the values
of $\sigma_0$ found for different temperatures are close to each
other that gives $\sigma_0=(7.3\pm 0.3)\,G_0$ and $k_Fl= 2.3\pm
0.1$. It should be stressed that $\beta=1$ whereas $\alpha<1$ in
accordance with Ref.~\onlinecite{Min04-2}.

Now, we return to  the saturation of the fitting parameter
$\tau_\phi^\ast$ with the decreasing temperature. Let us compare the
two characteristic lengths: localization length $\xi$ and the phase
breaking length, $L_\phi=\sqrt{D\tau_\phi}$, where $D=\pi
\hbar^2\sigma_0/(e^2m) $ is the diffusion coefficient. Using
Eq.~(\ref{eq00}) we have $\xi\simeq 40\,l$ for $k_Fl= 2.3$ found
above. The $L_\phi$ value can be estimated under the assumption that
the true phase breaking time $\tau_\phi$ is inversely related to the
temperature and  can be obtained by extrapolating of the
high-temperature data for $\tau_\phi^\ast$ as shown in
Fig.~\ref{F5}(b) by the dashed line. So, $\tau_\phi/\tau\simeq
1200/T $ for this case, that gives $L_\phi(T)\simeq 35\,l/T^{0.5}$.
Thus, $L_\phi$ should becomes comparable with $\xi$ at $T\sim 1$~K.
To realize the consequences, recall that for fully delocalized
electrons the increase of the magnitude of WL-correction with the
temperature decrease results from the fact that the area
contributing to the correction, which is proportional to $L_\phi^2$,
linearly increases. It is naturally to suppose that when the
electron states becomes partially localized, i.e., $\xi\sim L_\phi$,
the increasing of the interference correction at decreasing
temperature is restricted not only by $L_\phi^2$ but by $\xi^2$
also. Because of this,  the effective area, which contributes to the
interference correction, is $[(1/L_\phi)^2 +(1/\xi)^2]^{-1}$. In
time domain this corresponds to introduction of the  effective time
$\tau_\phi^\ast$:
\begin{equation}
\tau_\phi^\ast=\left(\frac{1}{\tau_\phi}+\frac{1}{\tau_\xi}\right)^{-1},
\label{eq50}
\end{equation}
where $\tau_\xi=\xi^2/D$. How may this fact reveal itself in
magnetoresistance? For large $k_Fl$-values, the shape of the
magnetoconductance curve is determined by competition between
$L_\phi$ and $l_H=\sqrt{\hbar/2eB}$. As for the considered case, the
area which contributes to WL is restricted by $[(1/L_\phi)^2
+(1/\xi)^2]^{-1}$ instead  of $L_\phi^2$ and  the shape of the
magnetoconductance curve will give $\tau_\phi^\ast$ instead of
$\tau_\phi$.

As evident from Fig.~\ref{F5}(b) this simple model well describes
the experimental $\tau_\phi^\ast$-vs-$T$ dependence within whole
temperature range if one supposes that $\tau_\xi/\tau=340$ and
$\tau_\phi/\tau=1200/T$ in Eq.(\ref{eq50}). The value of $\xi$ found
by this way, $\xi=\sqrt{D\tau_\xi}\simeq 19\,l$, somewhat differs
from that estimated from Eq.(\ref{eq00}), $\xi\simeq 40\,l$. Taking
into account the qualitative character of our consideration these
estimates can be viewed as consistent.

Thus, supposing that the conductivity is determined by delocalized
states, as for large $k_Fl$-values, and  the partial localization of
the states leads only to restriction of the WL correction  both by
$L_\phi$ and by $\xi$, we are able to describe quantitatively
 all the experimental data, namely, the magnetoresistance, the
temperature dependence of the conductivity at $B=0$, and the
temperature dependence of $\tau_\phi^\ast$.

\begin{figure}
\includegraphics[width=0.7\linewidth,clip=true]{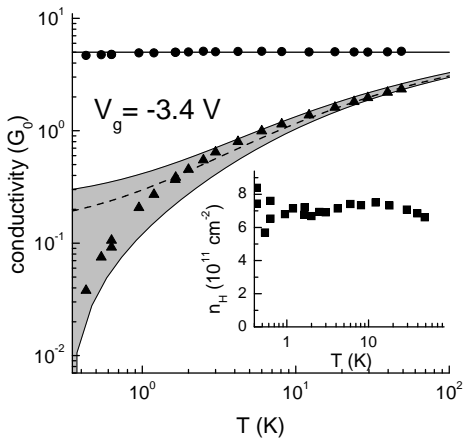}
\caption{ The conductivity at $B=0$ as a function of temperature for
$V_g=-3.4$~V. The values measured experimentally shown as triangles.
Circles are the Drude conductivity found experimentally for the
different temperatures as described in text. Solid line is
$\sigma_0=5.0\, G_0$. Dashed line is Eq.~(\ref{eq20}) with
$\tau_\phi^\ast/\tau$ shown in Fig.~\ref{F7}(b) by solid line.
Dashed area is uncertainty in calculated $\sigma(T)$ caused by
experimental error in determination of $\sigma_0$ and
$\tau_\phi^\ast$. The inset shows the temperature dependence of the
Hall electron density, $n_H=1/(eR_H)$. }
 \label{F6}
\end{figure}

\begin{figure}
\includegraphics[width=0.9\linewidth,clip=true]{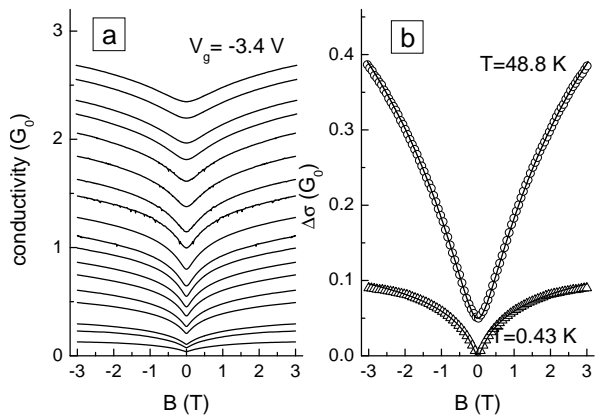}
\caption{(a) -- The magnetic-field dependence of the conductivity
measured at $V_g=-3.4$~V for different temperatures: $T=48.8$,
$39.3$, $29.5$, $24.0$, $18.0$, $12.2$, $8.1$, $6.0$, $4.2$, $3.0$,
$2.5$, $2.0$, $1.64$, $1.19$, $0.95$, $0.63$, $0.54$, and  $0.43$~K
(from the top to  bottom). (b) -- The $\Delta\sigma$-vs-$B$
dependences for $V_g=-3.4$~V, $T=48.8$~K and $0.43$~K. Symbols are
experimental data, lines are the fit by Eq.~(\ref{eq35}) carried out
within magnetic field range from $-0.3\, B_{tr}$ to $0.3\, B_{tr}$,
$B_{tr}=5.3$~T. For clarity, the data for $T=48.8$~K are shifted up
by  $0.05\,G_0$.}
 \label{F65}
\end{figure}

Let us analyze the data with farther increasing of  disorder.
Figs.~\ref{F6} and \ref{F65} show  the temperature and magnetic
dependences of the conductivity at $V_g=-3.4$~V, $n=6.5\times
10^{11}$~cm$^{-2}$. One can see that for this gate voltage the
conductivity at highest temperature is large enough
$\sigma(48.8\,\text{K})\simeq 2.4\, G_0$. It strongly decreases with
the lowering temperature and comprises only $0.038\, G_0$ at
$T=0.42$~K. Usually,  such a behavior  of the conductivity is
attributed with the crossover to the hopping regime. However, we
direct reader's attention to the Hall effect. The transverse
resistivity  $\rho_{xy}$ is linear with respect to magnetic field.
The Hall density $n_H=1/(eR_H)$ does not practically depend on the
temperature (see the inset in Fig.~\ref{F6}) and gives correct
density.  This seems extraordinary for the hopping regime wherein
the Hall effect is either absent or the value $1/(eR_H)$ has nothing
to do with the carrier
density.\cite{Friedman78-1,Friedman78-2,Look90,Nebel96}

\begin{figure}[b]
\includegraphics[width=0.9\linewidth,clip=true]{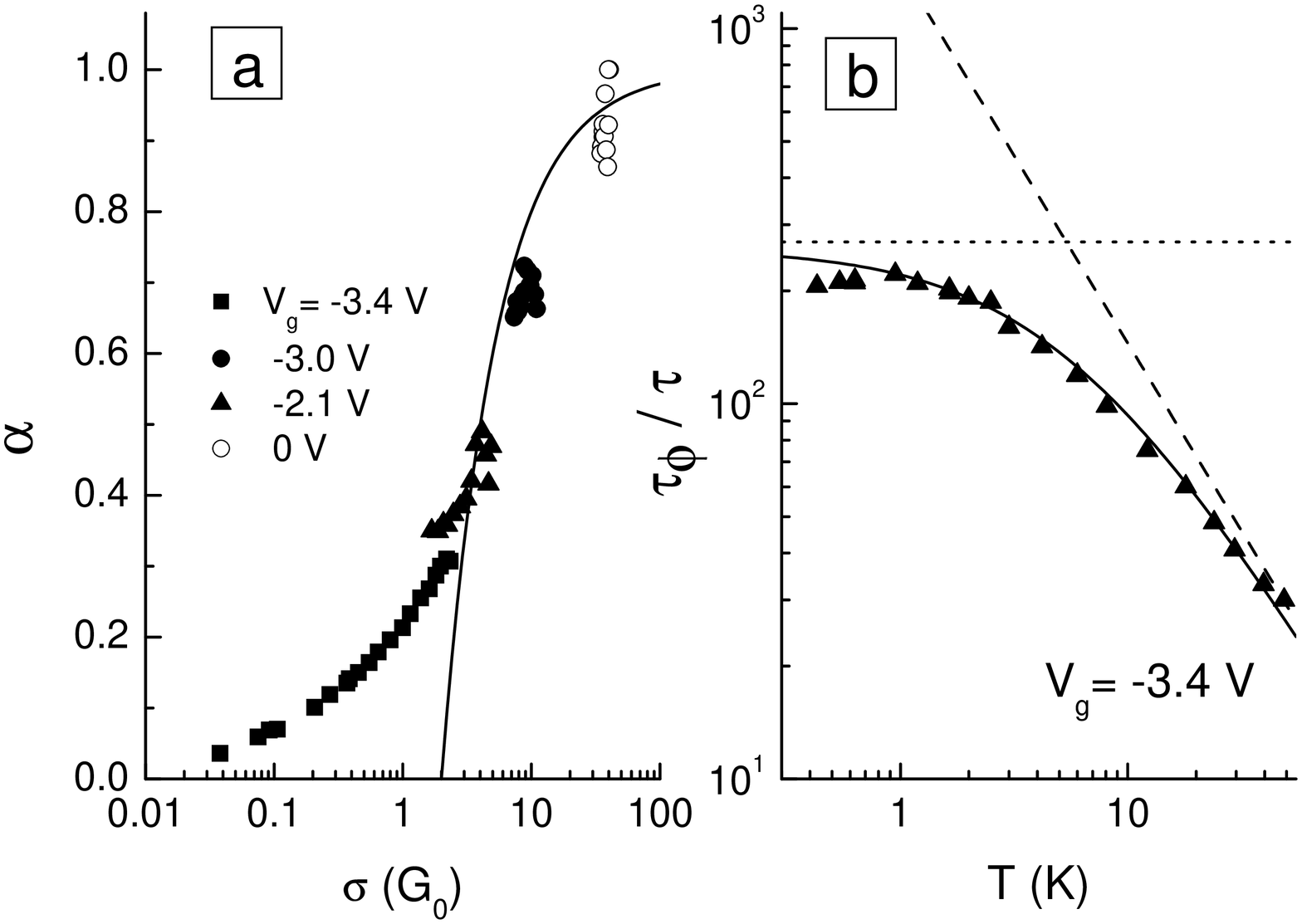}
\caption{(a) -- The prefactor $\alpha$ as a function of conductivity
for different  gate voltages. Symbols are the data, the solid line
is Eq.~(\ref{eq40}). (b) -- The temperature dependence of the
$\tau_\phi$ to $\tau$ ratio found experimentally for $V_g=-3.4$~V.
The dashed line is $\tau_\phi/\tau=1450/T$, the dotted line is
$\tau_\xi/\tau=260$, the solid line is
$\tau^\ast_\phi/\tau=1/(\tau/\tau_\phi+\tau/\tau_\xi)$.} \label{F7}
\end{figure}

So, the behavior of the Hall conductivity makes to believe that
delocalized states determine the conductivity yet. Therefore, let us
use  the HLN expression for description of the negative
magnetoresistance which is also observed  over the whole temperature
range [see Fig.~\ref{F65}(a)]. As in the above cases the
experimental data are well described by the HLN expression
[Fig.~\ref{F65}(b)]. The conductivity dependence of $\alpha$ and the
temperature dependences of $\tau_\phi^\ast$ are presented in
Figs.~\ref{F7}(a) and \ref{F7}(b), respectively. Again, if one
calculates the Drude conductivity from the data measured at
different temperatures using Eq.(\ref{eq20}) ($K_{ee}=0$ in this
case), we obtain the values,  which are practically independent of
the temperature (see Fig.~\ref{F6}). Thus, we have $\sigma_0\simeq
5.0\, G_0$ and $k_Fl\simeq 1.6$ in this case.

The temperature dependence of $\tau_\phi^\ast$ shown in
Fig.~\ref{F7}(b) is well described by Eq.~(\ref{eq50}) with
$\tau_\xi=260\,l$ that gives $\xi\simeq 16\,l$. Note that  this
value is close to that found from Eq.~(\ref{eq00}), $\xi\simeq
12\,l$.

\begin{figure}
\includegraphics[width=0.7\linewidth,clip=true]{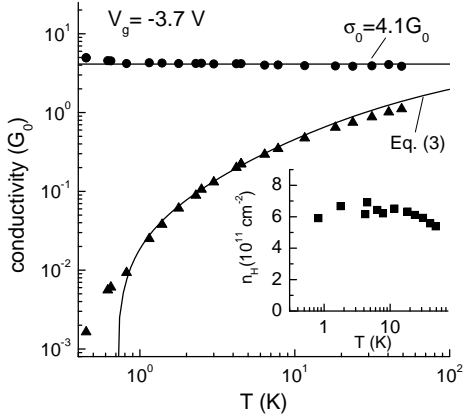}
\caption{The conductivity at $B=0$ as a function of temperature for
$V_g=-3.7$~V. The values measured experimentally shown as triangles.
Circles are the Drude conductivity found experimentally for the
different temperatures as described in text.  The dashed line is
calculated with $\Delta_\xi/\lambda=14$~K. The inset shows the
temperature dependence of the Hall electron density. }
 \label{F8}
\end{figure}

\begin{figure}[b]
\includegraphics[width=0.9\linewidth,clip=true]{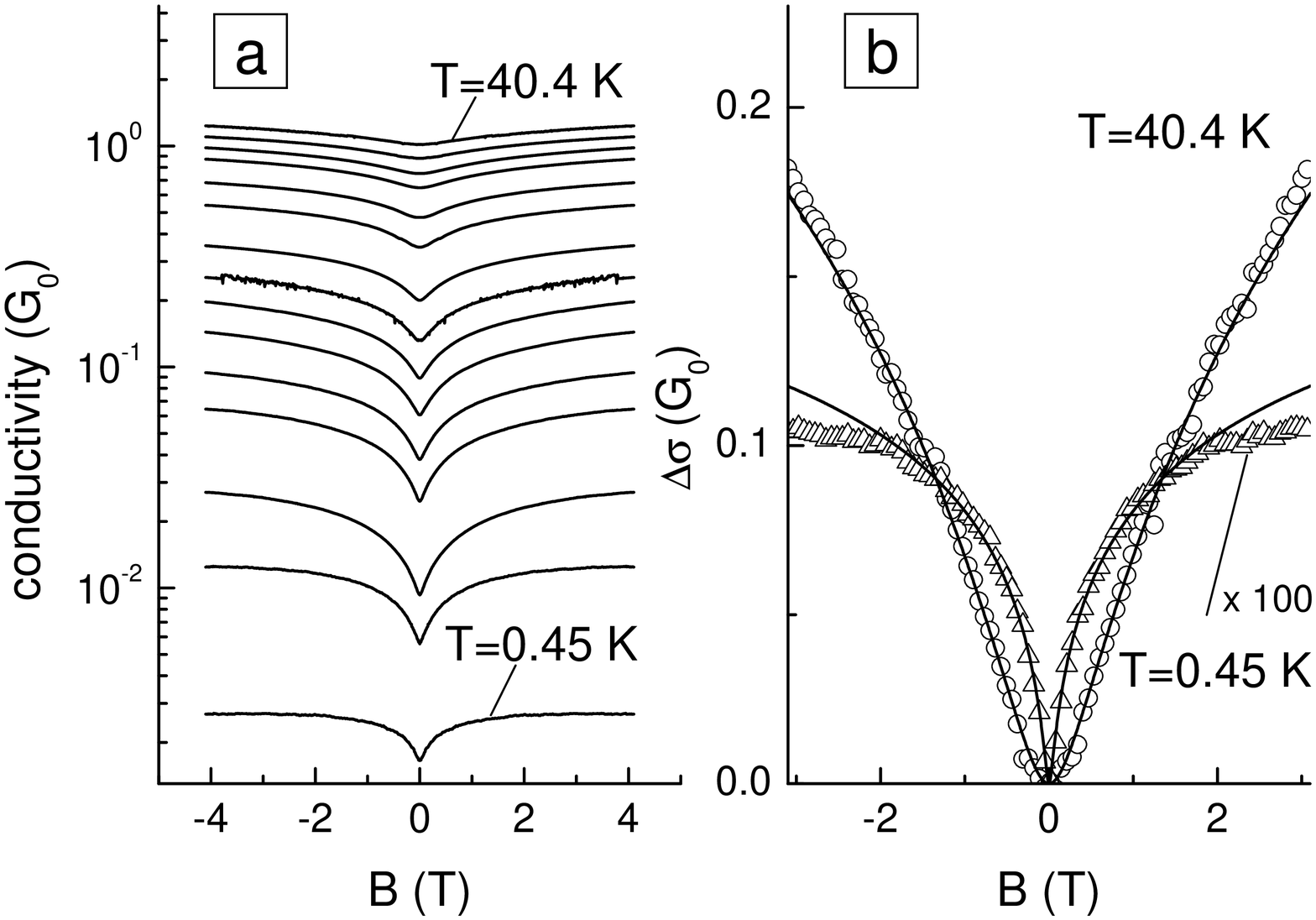}
\caption{ (a) -- The magnetic-field dependence of the conductivity
measured at $V_g=-3.7$~V for different temperatures: $T=40.4$,
$31.5$, $23.7$, $18.3$, $11.6$, $7.8$, $4.2$, $3.0$, $2.3$, $1.78$,
$1.39$, $1.15$, $0.82$, $0.62$, and $0.45$~K.  $B_{tr}=6.3$~T. (b)
-- The $\Delta\sigma$-vs-$B$ dependences for $V_g=-3.7$~V,
$T=40.4$~K and $0.45$~K. Symbols are experimental data, lines are
the fit by Eq.~(\ref{eq35}) carried out within magnetic field range
from $-0.3\, B_{tr}$ to $0.3\, B_{tr}$, $B_{tr}=6.3$~T.}
 \label{F85}
\end{figure}

\begin{figure}
\includegraphics[width=0.9\linewidth,clip=true]{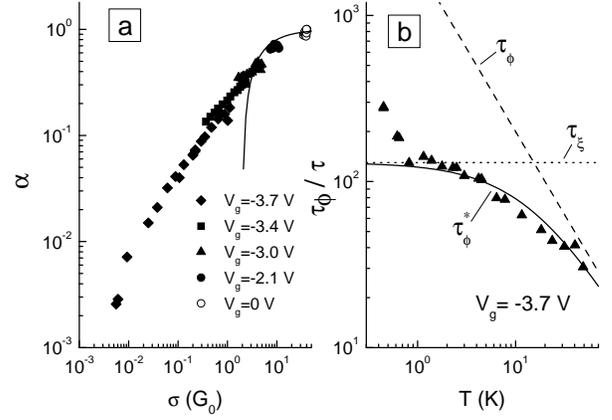}
\caption{(a) -- The prefactor $\alpha$ as a function of conductivity
for different  gate voltages. Symbols are the data, the solid line
is Eq.~(\ref{eq40}). (b) -- The temperature dependence of the
$\tau_\phi$ to $\tau$ ratio found experimentally for $V_g=-3.7$~V.
The dashed line is  $\tau_\phi/\tau=2000/T$, the dotted line is
$\tau_\xi/\tau=130$, the solid line is
$\tau^\ast_\phi/\tau=1/(\tau/\tau_\phi+\tau/\tau_\xi)$. } \label{F9}
\end{figure}

What does happen at farther decreasing of the electron
density?\footnote{The decreasing of electron density can lead to
that the Fermi level turns out below the classical mobility edge.
The study of the capacitance-voltage characteristics for our samples
shows that it is not the case.} The results for  $V_g=-3.7$~V,
$n=5.2\times 10^{11}$~cm$^{-2}$ are presented in Figs.~\ref{F8},
\ref{F85}, and \ref{F9}.

It is surprising, but the experimental temperature and magnetic
field dependences of the conductivity remains analogous to that in
the previous cases, the behavior of the fitting parameter of the
magnetoconductivity $\tau_\phi^\ast$ and $\alpha$ with temperature
correlates quite well also. The only difference is that
$\tau_\phi^\ast$ starts to grow sharply at $T \lesssim 0.8$~K
[Fig.~\ref{F9}(b)], when the conductivity becomes less than $10^{-2}
G_0$. This occurs when $\tau_\phi\simeq 20\, \tau_\xi$, i.e., the
dephasing length $L_\phi$ is about five times larger than the
localization length $\xi$. An important point is that the Hall
effect is also measurable down to this temperature (see inset in
Fig.~\ref{F8}). It is independent of the temperature, and $1/(eR_H)$
gives the electron density. It is impossible to measure the Hall
effect for $T\lesssim 0.8$~K. Fluctuations of the odd in $B$ voltage
across the Hall probes are significantly larger in magnitude than
the mean value of the voltage.

Thus, the analysis  of the experimental data within wide temperature
range, when both conditions $L_\phi<\xi$ and $L_\phi>\xi$ are
realized, shows that the simple model of the conductivity over
delocalized states  well describes all the data  when $\xi\gtrsim
0.2\, L_\phi$. This model reduces to the following: there is the
Drude conductivity; the main correction to it comes from the
interference; the value of the interference contribution is
$G_0\ln(\tau/\tau_\phi+\tau/\tau_\xi)$; the value of $\tau_\phi$ is
proportional to $1/T$ over the whole range of the conductivity; the
value of $\tau_\xi$ is $\sim\tau\exp{(\pi k_F l)}$; the
magnetoresistance over the whole  conductivity range is determined
by suppression of the interference in the magnetic field; the
$\sigma$-vs-$B$ curve has the same shape as for the high
conductivity, but it is controlled by
$(\tau/\tau_\phi+\tau/\tau_\xi)$ instead of $\tau/\tau_\phi$; the
magnetoresistance  has a prefactor that decreases with the lowering
conductivity.

The dashed curves in panels (a) in Figs.~\ref{F2}, \ref{F4},
\ref{F6}, and \ref{F8} show how this model describes $\sigma(T)$ at
$B=0$. The curves have been calculated with $\sigma_0$ found above
for each concrete case and with $1/\tau_\phi^\ast(T)$ shown by solid
lines in panels (b) in Figs.~\ref{F3}, \ref{F5}, \ref{F7}, and
\ref{F9}. One can see that down to $\sigma\sim 0.01\,G_0$ this
simple model well describes the experimental data.

Thus, at low-temperature conductivity within the range
$(10^{-1}-10^{-2})\,G_0$ which occurs at the Drude  conductivity
about $(4-5)\,G_0$ all the transport properties look like ones for
the conductivity over delocalized states. However, it is clear that
for the case when the interference correction becomes comparable
with the Drude conductivity, the description of the system starting
from ``pure'' states of the conduction band is inappropriate (see
Fig.~\ref{F6}). It seems that another model of the charge transfer
could be more suitable for this case.   Namely, an electron can be
considered as being localized within the area $\sim \xi^2$ during
the time $\tau_\phi$. After this time, it leaves this area shifting
by a some length $L_i$. Analogous model was considered in
Refs.~\onlinecite{Gog75,Thouless77,Gog83,Gor05}. It is natural to
suppose that the length $L_i$ should be shorter than the
localization length $\xi$ and longer than the mean distance between
scatterers, $N_d^{-1/2}$: $\xi\gtrsim L_i\gtrsim N_d^{-1/2}$. Such a
motion looks like the motion of delocalized electron with the
diffusion coefficient $D\simeq L_i^2/\tau_\phi$ yielding the
conductivity
\begin{equation} \sigma(T)\simeq
e^2\nu\frac{L_i^2}{\tau_\phi(T)}
=\frac{L_i^2}{l^2}\,\frac{\tau}{\tau_\phi(T)}\,\sigma_0,
\label{eq75}
\end{equation}
where $\nu$ is density of the states.

\begin{figure}
\includegraphics[width=0.7\linewidth,clip=true]{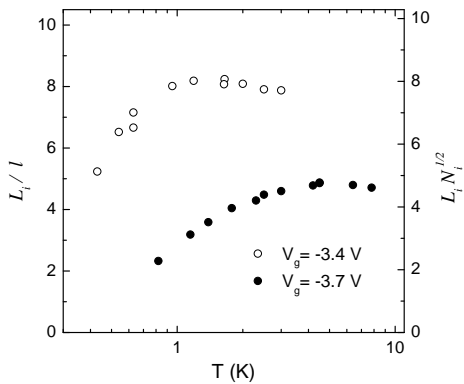}
\caption{The $L_i$ to $l$ ratio as a function of temperature for two
gate voltages.} \label{F11}
\end{figure}

To understand the soundness of the model, let us estimate the
shifting length $L_i$. Using $\tau_\phi(T)/\tau$ shown for
$V_g=-3.4$~V and $-3.7$~V by the dashed lines in Figs.~\ref{F7}(b)
and \ref{F9}(b), respectively,  and the corresponding experimental
$T$-dependences of $\sigma$ from Figs.~\ref{F6} and \ref{F8}, we
have found the length $L_i$ as a function of temperature from
Eq.~(\ref{eq75}) (see Fig.~\ref{F11}). One can see that the values
of $L_i$ are reasonable. They lie within the interval from $L_i
\simeq 2\,N_d^{-1/2}\simeq 0.3 \xi$ at $\sigma\simeq 0.01\,G_0$
(that corresponds to $V_g=-3.7$~V and $T=0.8$~K) up to $L_i\simeq
8\,N_d^{-1/2}\simeq 0.5\,\xi$ at $\sigma\simeq 0.5\, G_0$ (that
corresponds to $V_g=-3.4$~V and $T\simeq 1$~K). It seems natural
that the Hall effect for this conductivity mechanism should exist
and give the electron density. Unfortunately, this model was not
elaborated up to now and it makes no prediction about the magnetic
field dependences of the conductivity or other effects.

\section{Analysis from alternative standpoints}

The conductivity of 2D systems is usually interpreted  as the
hopping one when it is less than $e^2/h$. Such consideration rests
on a strong temperature dependence of the conductivity observed
under this condition. This dependence for the different hopping
mechanisms obeys the law
\begin{equation}
\sigma(T)=\sigma^0\exp\left[-\left(\frac{T_0}{T}\right)^p\,\right],
\,\,\,  \label{eq70}
\end{equation}
where $p=1$ for the nearest neighbor hopping, $p=0.5$ or $0.3$ for
the variable range hopping (VRH) with the Coulomb gap or without it,
respectively.

Additional argument is that the mean level spacing in localization
length $\Delta_\xi=1/(\nu\xi^2)$ becomes large than $T$. Really, the
strong temperature dependence is observed in our case when
$\sigma<e^2/h$ (see Figs.~\ref{F6} and \ref{F8}). The estimate of
the mean level spacing, for example, for $V_g=-3.4$~V gives
$\Delta_\xi\simeq 0.3-0.5$~meV $\simeq 4-6$~K. From the aforesaid,
it can mean that the probability of transition from one localized
state  to another becomes small already at $T\lesssim 4-6$~K, the
states are well localized, and the conductivity  is hopping.
However, this estimation  is very crude. The value $\Delta_\xi$ is
the mean level spacing {\it within} the localization length, while
the mean level spacing {\it between} the neighbor localized states,
$\Delta_\xi^n$, is of importance for the conductivity. Taking into
account the fact that a localized state has a number of neighbors
one can conclude that $\Delta_\xi^n$ should be several times less
than $\Delta_\xi$. Thus, we believe that the crossover to the
hopping in the case considered has to be observed at the
temperature, which is several times lower than $\Delta_\xi$.

\begin{figure}
\includegraphics[width=0.8\linewidth,clip=true]{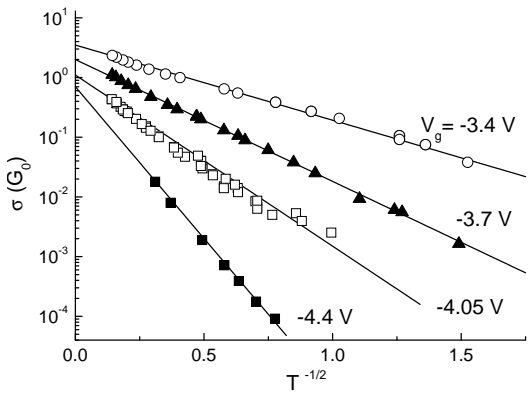}
\caption{The temperature dependence of the conductivity for
different gate voltages plotted in the VRH coordinates. Symbols are
the experimental results, lines are provided as a guide for the
eye.} \label{F10}
\end{figure}

Let us inspect our data from the ``hopping'' point of view.
Figure~\ref{F10} shows the temperature dependences of the
conductivity at the different gate voltages $V_g\leqslant-3.4$~V
when the conductivity  becomes less than $e^2/h$. One can see that
the data are well linearized in the coordinates
$\ln\,\sigma$-vs-$T^{-p}$ with $p=0.5$ within wide range of the
temperature and conductivity. This were meant that one deals with
the variable range hopping with the Coulomb gap.

Note, the most of the  published data are also linearized in the VRH
coordinates at $\sigma<e^2/h$. Practically all the authors present
this fact as the main argument in favor the hopping mechanism of
conductivity.\cite{Keus97-1,Keus97,Shlim00,Gersh00,Camin03} Very
wide range of the temperature and conductivity values, where the VRH
law is observed, is reckoned as important argument {\it pro}. From
our point of view it is argument {\it contra} rather than {\it pro}.
Really, the equality  $p=0.5$ for our case might mean that the
variable range hopping with the Coulomb gap is the main conductivity
mechanism over the whole temperature range from $0.4$~K up to
$60$~K. However, such a statement looks strange. Physically, it
seems natural that  the following conductivity regimes should change
each other with the increasing temperature: the VRH with Coulomb gap
($p=1/2$) $\rightarrow$  the Mott law ($p=1/3$) $\rightarrow$ the
nearest-neighbor hopping with $p=1$ (the regime of so called
$\varepsilon_3$-conductivity). Finally,  the transition (or
crossover) to the conductivity over the delocalized states should
happen. It seems that the variation of the temperature from
$T\approx 0.04$~meV up to $T\approx 5$~meV should cover all these
regimes, and the change both of the power $p$ and of the
characteristic  temperature $T_0$ in the law given by
Eq.~(\ref{eq70}) has to be observed experimentally.

The next strangeness  of such ``hopping'' conductivity is the fact
that  the conductivity  is well extrapolated at $T\to \infty$ to the
value $\sigma^0$, which is close to $e^2/h$. It is the value of the
conductivity of one open channel. It seems that the conductivity
between two localized states should be significantly lower. To
interpret the high value of $\sigma^0$, the new hopping mechanism,
so called the electron-electron interaction assistant hopping, was
proposed in Refs.~\onlinecite{Kozub00} and \onlinecite{Shlimak01}.
For the best of our knowledge there is not consistent theory for
this hopping mechanism. In fact, it was found already in
Ref.~\onlinecite{Flei80}  that in the absence of phonons, the {\em
e-e} interaction is not sufficient to support the variable range
hopping in dimensionality $d<3$, even for the case of long-range
Coulomb interaction. This result was recently confirmed in Refs.
\onlinecite{Gor05} and \onlinecite{Basko05}.

Third, as already mentioned above  the Hall effect is not measurable
or the quantity $1/(eR_H)$ is not equal to the electron density for
the case of conventional hopping
conductivity.\cite{Friedman78-1,Friedman78-2,Look90,Nebel96} The
insets in Figs.~\ref{F6} and \ref{F8} show that the Hall coefficient
in our experiment is independent of temperature, and
Fig.~\ref{F1}(a) demonstrates that $1/(eR_H)$ gives the electron
density.

All these  features become transparent if one analyzes  the data
down to $\sigma\sim 10^{-2} G_0$ within the framework of the
conductivity over delocalized states. First, the temperature
dependences of the conductivity are consistently described  within
whole temperature range.

The second, the extrapolation of  $\sigma(T)$ to $e^2/h$ at
$T\to\infty$ is absolutely natural because the strong temperature
dependence of the conductivity is observed when the Drude
conductivity is close to $e^2/h$. Therefore, independently of the
manner,  $\sigma(T)$  has to be extrapolated to the value close to
$e^2/h$.

At last, the existence  of the Hall effect, which gives the
electron density, is natural for the conductivity over the
delocalized states.

Thus, we believe that the hopping is inappropriate model for the
description of the conductivity within the range $(1-10^{-2})\,G_0$.

\begin{figure}
\includegraphics[width=0.9\linewidth,clip=true]{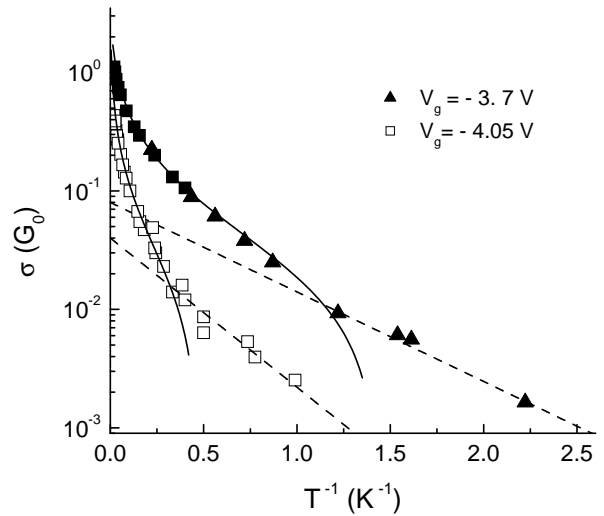}
\caption{The  temperature dependence of the conductivity for two
gate voltages, illustrating the crossover to the
$\varepsilon_3$-conductivity. Symbols are the experimental data,
solid lines are drown according to the model considered in
Section~\ref{sec:res} with $\sigma_0=4.1\,G_0$,
$\tau_\phi/\tau=2000/T$, $\tau_\xi/\tau=130$  for $V_g=-3.7$~V and
$\sigma_0=3.4\,G_0$, $\tau_\phi/\tau=3800/T$,  $\tau_\xi/\tau=62$
for $V_g=-4.05$~V. The dashed lines are Eq.~(\ref{eq110}) with
$\sigma_3=0.08\, G_0$, $\varepsilon_3=0.15$~meV for $V_g=-3.7$~V and
$\sigma_3=0.04\, G_0$, $\varepsilon_3=0.25$~meV for $V_g=-4.05$~V.}
\label{F115}
\end{figure}

It may appear from reading of this and our previous papers that we
deny the hopping conductivity mechanism in the 2D systems under
investigation. This is not the case. As mentioned above the hopping
between the nearest localized states, to our opinion, becomes to be
the main conductivity mechanism when the conductivity becomes lower
$\sim 10^{-2}\,G_0$. In this case the $T$-dependence of the
conductivity should follow the law
\begin{equation}
\sigma(T)\simeq \sigma_3
\exp{\left(-\frac{\varepsilon_3}{T}\right)}. \label{eq110}
\end{equation}
The experimental data for $V_g=-3.7$~V and $-4.05$~V are represented
in Fig.~\ref{F115} as $\lg\sigma$-vs-$T^{-1}$ plots in accordance
with this point. One can really see that they approach exponential
law, Eq.~(\ref{eq110}), with the decreasing temperature. As this
takes place, the conductivity  is less than $10^{-2}G_0$, therewith
the value of $\sigma_3$ is significantly lower than $e^2/h$. The
experimental activation energy $\varepsilon_3$ is $0.15$~meV and
$0.25$~meV for the cases presented in Fig.~\ref{F115}. Let us
compare these values  with the mean level spacing $\Delta_\xi$. The
estimate for $V_g=-3.7$~V gives $\Delta_\xi\simeq 0.5$~meV and
$\simeq 1$~meV depending on that what value of $\xi$,
$\sqrt{D\tau_\xi}$ or $l\,\exp{(\pi k_Fl/2)}$, we are using. For
$V_g=-4.05$~V we, respectively, obtain $\Delta_\xi\simeq 0.7$~meV
and $\simeq 1.5$~meV. The fact that the $\varepsilon_3$-energy
occurs three-seven times less than $\Delta_\xi$ is not surprising.
As discussed in the beginning of this section the mean level spacing
between the neighbor localized states $\Delta_\xi^n$ is of
importance for the conductivity, rather than the mean level spacing
within the localization length $\Delta_\xi$ which is essentially
larger.

In Fig.~\ref{F115}, we present also the temperature dependences of
the conductivity  calculated within the framework of the model
described in Section~\ref{sec:res}. It is seen that the hopping
regime  matches the regime of the conductivity over the delocalized
states quite well. The crossover happens when $\sigma\sim 10^{-2}\,
G_0$.

All the above analysis was carried out within the model of
single-particle localization. The {\em e-e} interaction was taken
into account through the phase-breaking time only. Recently, the
role of the many-body localization in the transport properties of
low-dimensional systems was discussed in
Refs.~\onlinecite{Gor05cmv1}, \onlinecite{Gor05},
\onlinecite{Basko05}, and \onlinecite{Basko06cm}. It was shown  that
effects of Anderson localization in the many-body Fock space become
efficient and dramatically suppress the direct current  conductivity
for $T\alt \Delta_\xi/\lambda$, where $\lambda^2\ll 1$ is
dimensionless coupling characterizing the {\em e-e} interaction,
which is assumed to be short-range and weak. In particular, the
many-body localization makes the variable-range hopping transport
mechanism impossible in 2D in the absence of phonons.\cite{Flei80}
The direct current conductivity should exactly vanish at finite low
temperatures, in accordance with the prediction of
Ref.~\onlinecite{Flei80}; the corresponding characteristic critical
temperature is given by $T_c\sim \Delta_\xi/(\lambda|\ln\lambda|).$
In Ref.~\onlinecite{Basko05} the stability of the metallic phase for
$T\gg T_c$ and the insulating phase for $T\ll T_c$ was proved;  the
critical $T$-dependence of the conductivity in the vicinity of $T_c$
was proposed in Ref.~\onlinecite{Gor05}.

Since it is the value of $\lambda^2$ that  determines the
$\tau_\phi$-vs-$T$ dependence, $\tau_\phi^{-1}\sim \lambda^2\, T$,
it can by easily estimated from the experimental data. For instance,
we obtain $\lambda\simeq 0.4$ for $V_g=-3.7$~V using $\tau_\phi(T)$
shown   in Fig.~\ref{F9}(b). With this value and
$\Delta_\xi=0.5-1$~meV, we have $T_c\simeq 15-30$~K (the very close
results are obtained for other gate voltages). As evident from the
above we observe no evidence of the critical behavior of $\sigma$ at
these temperatures as well as in the entire  temperature range.
(Actually, the expression for $T_c$ may in principle contain
numerical factors which would decrease the value of $T_c$ by an
order in magnitude as compared to the above estimate, see discussion
in the beginning of this section).

Resuming this section we would like to underline that it is
insufficient to analyze only the temperature dependence of the
conductivity in order to establish  the conductivity mechanism. It
is necessary to know other theoretical predictions concerning, for
example, the Hall effect, behavior of the conductivity in the
magnetic and electric fields.

\section{Conclusion}
We have studied the temperature and magnetic field dependences of
the conductivity in heavily doped, strongly disordered 2D quantum
well structures within wide conductivity and temperature ranges. The
role of interference in the temperature and magnetic field
dependences has been traced when the phase breaking length
controlled by the temperature occurs to be both larger and smaller
than the localization length.

\begin{figure}
\includegraphics[width=\linewidth,clip=true]{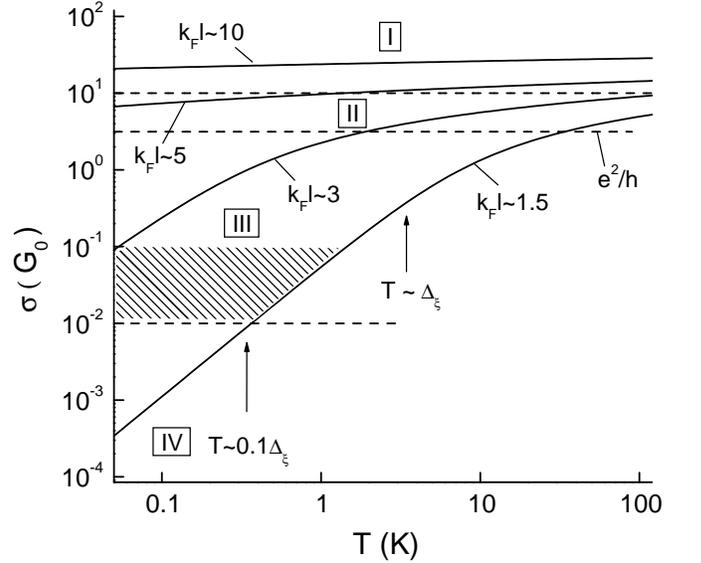}
\caption{The conductivity regimes in two-dimensions. Solid lines are
the balk-park temperature dependences of 2D disordered system with
different $k_Fl$-values. The dashed lines conventionally demarcate
the different areas (see text).} \label{F12}
\end{figure}

Our conclusions are pictorially summarized in Fig.~\ref{F12} and
accumulated in Table~\ref{tab1}. It has been shown that the four
different areas, labeled as  I -- IV, can be fictionally
distinguished in the $T$-dependence of the conductivity of 2D
systems with different disorder strength. The boundaries between the
areas are relatively smooth and shown in Fig.~\ref{F12}
approximately.

At $k_Fl\gtrsim 4-5$ all the temperature and magnetic field
dependences are in a good agreement with the conventional theories
of the quantum corrections. This is area I. Here, inequality $\xi\gg
L_\phi$ is fulfilled at any accessible temperatures.

At $k_Fl\lesssim 4$ and $\sigma\gtrsim (2.5-3)\, G_0$, that
corresponds to area II,  the magnitude of the interaction correction
tends to zero when $k_Fl$ lowers. The low-magnetic-field
magnetoconductivity caused by suppression of the interference
quantum correction is described  by the HLN-expression,
Eq.~(\ref{eq35}), with the prefactor $\alpha$, $\alpha \simeq
1-2G_0/\sigma$.

In the area III, the localization length $\xi$ becomes comparable
with the dephasing length $L_\phi$. The interference contribution to
the conductivity is now restricted not only by the phase breaking
length $L_\phi$, but by the localization length $\xi$ as well. The
quantity $(\tau_\phi^\ast)^{-1}=\tau_\phi^{-1}+\tau_\xi^{-1}$ rather
than $\tau_\phi^{-1}$ determines  the behavior of the conductivity,
$\tau_\phi\propto T^{-1}$ and $\tau_\xi\sim\tau\exp(\pi k_F l)$
therewith. The temperature dependence of the conductivity within
this range can be well described by the formula
$\sigma\simeq\sigma_0+G_0\,\ln{(\tau/\tau_\phi^\ast)}$  down to
$\sigma\sim 10^{-2}\, G_0$. The magnetoconductivity is caused by
suppression of the quantum interference. The $\sigma$-vs-$B$
experimental plots are well described by the same formula, as in the
weak-localization regime,  Eq.~(\ref{eq35}), with $\tau_\phi^\ast$
instead of $\tau_\phi$ and with the prefactor, which value decreases
with the lowering conductivity. This regime is realized in our
temperature range  when $k_Fl\lesssim 2-3$ and $L_\phi\simeq
(0.5-4)\,\xi$.

In the part of the area III shown by hatching in Fig.~\ref{F12}, in
which $\sigma\sim (10^{-2}-10^{-1})\, G_0$, the other model
considering the transport as the diffusion motion with $D\sim
L_i^2/\tau_\phi$ ( $\xi \gtrsim L_i\gtrsim N_d^{-1/2}$) could be
more appropriate for description of the transport.

\begin{table}
\caption{The $T$- and $B$-dependences of the conductivity for
different areas shown in Fig.~\ref{F12}} \label{tab1}
\begin{ruledtabular}
\begin{tabular}{ccc}
Area & $\sigma(T,B=0)$ & $\Delta\sigma(B)$\\
 \colrule
I& $\sigma_0+\ln(\tau/\tau_\phi)+K_{ee}\ln{T\tau}$ &
$\alpha\, G_0\, {\cal H}(\tau/\tau_\phi,B/B_{tr})$,\footnotemark[1]  \\
&&$\alpha=1$\\[2mm]

 II& $\sigma_0+\ln(\tau/\tau_\phi)+K_{ee}\ln{T\tau}$ & $\alpha\,
G_0\,
{\cal H}(\tau/\tau_\phi,B/B_{tr})$,  \\
&  $K_{ee}\to 0$ @ $k_Fl \searrow$ & $\alpha=1-2/\sigma$\\[2mm]
 III& $\sigma_0+\ln(\tau/\tau_\phi+\tau/\tau_\xi)$ &
$\alpha\, G_0\, {\cal H}(\tau/\tau_\phi+\tau/\tau_\xi,B/B_{tr})$,\,\,\\
& $\tau_\xi\sim \tau\exp{(\pi k_F l)}$ &$\alpha\to 0$ @ $\sigma \searrow$ \\[2mm]
IV&$\sigma_3\exp{(-\varepsilon_3/T)}$ & determined by $\xi(B)$  \\
 & $\varepsilon_3\simeq (0.1-0.3)\,\Delta_\xi$& \\
 \end{tabular}
\end{ruledtabular}
\footnotetext[1]{The function ${\cal H}$ is defined in
Eq.~(\ref{eq35}).}
\end{table}

At last, when the temperature becomes $5-10$ times smaller than the
mean level spacing in localization length $\Delta_\xi$, the hopping
between nearest neighbors becomes the main conductivity mechanism
(area IV). For our case it occurs when the conductivity is less than
$\sim 10^{-2}\,G_0$.

To our opinion, the scenario described is, in general, common for
all disordered systems with two-dimensional gas of weak-interacting
electrons ($r_s\lesssim 1$), in which spin-dependent processes are
unimportant. Only the details, such as the temperature scale and the
positions of the boundaries between areas,  may be different for
each concrete case.

\subsection*{Acknowledgment}
We are grateful to Igor Gornyi for very useful discussions. This
work was supported in part by the RFBR (Grant Nos. 04-02-16626,
05-02-16413, and 06-02-16292), and by a Grand from the  President of
Russian Federation for Young Scientists MK-1778.2205.2.


\end{document}